\documentclass{article}

\usepackage{microtype}
\usepackage{graphicx}
\usepackage{subfigure}
\usepackage{subcaption}
\usepackage{booktabs} 
\usepackage{multirow}
\usepackage{colortbl}
\usepackage{xcolor}
\usepackage{arydshln}
\usepackage{hyperref}

\usepackage[preprint]{icml2026}
\usepackage{bbm}
\usepackage{amsmath}
\usepackage{amssymb}
\usepackage{mathtools}
\usepackage{amsthm}
\usepackage{booktabs}       
\usepackage{threeparttable} 
\usepackage{makecell}       

\usepackage[capitalize,noabbrev]{cleveref}

\theoremstyle{plain}

\theoremstyle{definition}

\theoremstyle{remark}

\usepackage{tikz}

\usepackage[textsize=tiny]{todonotes}

\usepackage{graphicx} 
\usepackage{booktabs} 

\icmltitlerunning{Submission and Formatting Instructions for ICML 2026}

\begin{document}

\twocolumn[
  \icmltitle{Privacy-Preserving LLMs Routing}



  \icmlsetsymbol{equal}{*}
\icmlsetsymbol{corresponding}{$\dagger$}
  \begin{icmlauthorlist}
    \icmlauthor{Xidong Wu}{equal,sch1}
    \icmlauthor{Yukuan Zhang}{equal,sch2}
    \icmlauthor{Yuqiong Ji}{sch3}
    \icmlauthor{Reza Shirkavand}{sch5}
    \icmlauthor{Qian Lou}{corresponding,sch2}
    \icmlauthor{Shangqian Gao}{corresponding,sch4}
  \end{icmlauthorlist}

  \icmlaffiliation{sch1}{Department of Electrical and Computer Engineering, University of Pittsburgh}
  \icmlaffiliation{sch2}{Department of Computer Science, University of Central Florida}
  \icmlaffiliation{sch3}{Department of Computer and Information Science, University of Pennsylvania}
  \icmlaffiliation{sch5}{Department of Computer Science, University of Maryland}
  \icmlaffiliation{sch4}{Department of Computer Science, Florida State University}

\icmlcorrespondingauthor{Qian Lou}{qian.lou@ucf.edu}
\icmlcorrespondingauthor{Shangqian Gao}{sgao@cs.fsu.edu}
  \icmlkeywords{Machine Learning, ICML}

  \vskip 0.3in
]



\printAffiliationsAndNotice{\icmlEqualContribution}

\begin{abstract}
Large language model (LLM) routing has emerged as a critical strategy to balance model performance and cost-efficiency by dynamically selecting services from various model providers. However, LLM routing adds an intermediate layer between users and LLMs, creating new privacy risks to user data. These privacy risks have not been systematically studied.
Although cryptographic techniques such as Secure Multi-Party Computation (MPC) enable privacy-preserving computation, their protocol design and implementation remain under-explored, and na\"ive implementations typically incur prohibitive computational overhead. To address this, we propose a privacy-preserving LLM routing framework (PPRoute). PPRoute includes multiple strategies to speed up encoder inference and nearest neighbor search under the MPC and maintain the quality of LLM routing. First, PPRoute uses MPC-friendly operations to boost the encoder inference. Second, PPRoute uses a multiple-step model training algorithm to maintain routing quality despite the constraints of the encrypted domain. Third, PPRoute proposes an unsorted Top-k algorithm with $O(1)$ communication complexity for secure sorting in model search, significantly reducing communication latency.  Across different datasets, PPRoute achieves the performance of plaintext counterparts, while achieving approximately a 20$\times$ speedup over na\"ive MPC implementations.


\end{abstract}

\section{Introduction}

Recent advances in large language models (LLMs) show impressive performance across diverse natural language tasks, driven by the large model size and extensive training on massive datasets. Despite these gains, a trade-off remains between the sophisticated reasoning of frontier models and the cost-efficiency of their smaller counterparts. This disparity creates a deployment challenge: routing queries to the most capable model is prohibitively costly, whereas relying exclusively on smaller models compromises output quality. 

To address the inherent trade-off between model capability and inference overhead, 
LLM routing is designed to assign simpler prompts toward lightweight models and reserve high-parameter models for complex reasoning \cite{chen2023frugalgpt,wang2023tabi}. By combining open-source and closed-source models and integrating services from different model providers, LLM routing can dynamically direct incoming queries to the most suitable model. Many LLM routing algorithms \cite{yue2023large,ong2025routellm} have been developed and adopted. However, this architectural layer introduces significant security vulnerabilities because (1) the integration of a routing stage adds potential data leakage within the inference pipeline.
(2) LLM routing is typically managed by a third party rather than by the primary model providers, complicating the chain of control for sensitive data; and (3) currently, privacy-preserving methodologies for LLM routing remain largely unexplored.

Secure Multi-Party Computation (MPC) \cite{evans2018pragmatic} is a cryptographic technique that coordinates multiple parties to jointly compute functions over their private data. By operating on secret-shared data, MPC ensures that input data, intermediate values, function parameters, and final outputs remain confidential throughout the computation. In the context of machine learning, MPC enables privacy-preserving inference by securely evaluating neural networks without exposing user queries or learned parameters \cite{evans2018pragmatic}. However, 
directly applying MPC to LLM routing incurs prohibitive computational overhead. Secure arithmetic and communication-heavy protocols significantly increase latency, particularly for non-linear components such as activation functions and softmax \cite{mohassel2017secureml}. For example, the $BERT_{\text{BASE}}$ inference under MPC can be more than $60\times$ slower than its plaintext counterpart \cite{li2023mpcformer}. Consequently, designing LLM routing mechanisms that leverage cryptographic techniques to provide strong privacy protection while maintaining acceptable runtime and comparable accuracy remains an open research problem.

In this paper, we take the first step toward privacy-preserving LLM routing under MPC without sacrificing routing effectiveness. We design a new routing framework built on embedding-based LLM routing because embedding-based LLM routing is model-agnostic, maintains low inference overhead, and generalizes effectively to both unseen LLMs and out-of-distribution prompts. Embedding-based LLM routing maps both models (offline) and user queries (online) into a shared high-dimensional vector space. The core objective is to route requests to the most efficient model by balancing semantic relevance and cost constraint, utilizing either clustering or K-Nearest Neighbors (KNN) \cite{jitkrittum2025universal,shirkavand2025cost}.
While efficient in plaintext environments, implementing these routers under MPC introduces significant overhead: 1) Running an encoder model inside MPC involves complex computation and communication, leading to high latency during the real-time inference phase; 2) The nearest neighbor search in model selections requires heavy comparison and sorting operations.

To address these two challenges, we propose Privacy-Preserving LLMs Routing (PPRoute) to optimize encoder inference and nearest neighbor search, which can be directly applied to embedding-based LLM routing. The PPRoute is shown in \Cref{pproute} with CSCR as an example. In summary, the main contributions of this paper are as follows:

\begin{figure*}[!t]
  \begin{center}
    \centerline{\includegraphics[width=1.7\columnwidth]{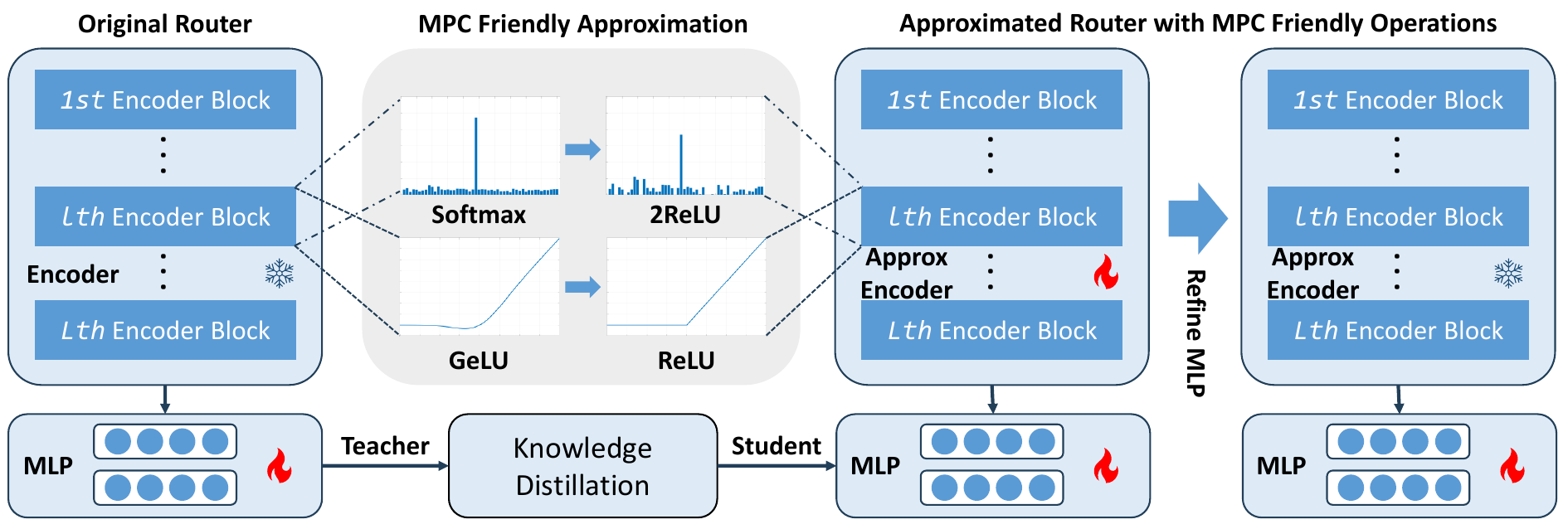}}
    \caption{
     Overview of Privacy-Preserving LLMs Routing (PPRoute) Training. 1) Train the MLP layer in the teacher model. 2) The student model uses the MPC-friendly operations in the approximated encoder, and model training follows \cref{eq:loss}. 3) Train the MLP layers only and encrypt the model to serve under the MPC.
    }
    \label{pproute}
  \end{center}
\end{figure*}


\begin{itemize}
\item To the best of our knowledge, this is the first work to systematically investigate the privacy risks in LLM routing and propose an end-to-end privacy-preserving LLMs Routing (PPRoute) framework to protect user queries in inference. It achieves comparable quality  to the unencrypted counterpart, with only a reasonable runtime overhead.

\item 
PPRoute optimizes encoder inference efficiency by removing computationally expensive operations. By replacing Softmax and ReLU with 2ReLU and ReLU, respectively, the framework ensures a faster, more secure inference process.

\item PPRoute utilizes a multi-stage distillation algorithm to maintain high routing accuracy while significantly lowering inference overhead. 

\item PPRoute introduces a novel constant-round unsorted Top-k algorithm with $O(1)$ communication complexity for secure retrieval in LLM routing. By reducing the number of communication rounds to a constant, it significantly accelerates secure nearest neighbor search.
\end{itemize}

\section{Preliminary and Related Work}

\subsection{LLM router}

LLM routing dynamically selects among multiple large language models to serve a given query to balance cost, latency, and output quality \cite{woisetschlager2025mess, huang2025lookahead}. 
LLM-Blender \cite{jiang2023llm} proposes an ensemble framework that queries multiple LLMs at inference time and selects the best response to improve output quality. Cascade-based methods are proposed to route queries through cheaper or smaller models first and defer only difficult cases to larger models, significantly reducing cost without sacrificing performance \cite{chen2023frugalgpt,yue2023large}. Several systems formalize this idea with learned routing mechanisms, using calibrated confidence scores, predicted query difficulty, or meta-models that estimate LLM performance to decide whether a small model is sufficient or escalation is required \cite{wang2023tabi,vsakota2024fly,ding2024hybrid}. Extra signals, such as reward signal \cite{lu2024routing, zhang2025router}, human preference data \cite{ong2025routellm,tsiourvas2025causal}, dynamic verification signals from smaller models \cite{aggarwal2023automix, wang2025reason}, or graph neural networks with edge prediction mechanisms \cite{feng2024graphrouter}, can also be used to train LLM routers, allowing more precise assignment of queries to models with appropriate expertise. 

In addition, embedding-based LLM routing methods have been proposed \cite{chen2024routerdc,zhuang2024embedllm}, which enable model-agnostic routing by seamlessly integrating closed-source or third-party models, incur low inference overhead since routing decisions rely on compact embeddings rather than full model inference, and exhibit robust generalization to unseen LLMs and out-of-distribution prompts. 
They leverage query similarity to guide model selection, providing a lightweight alternative to fully learned routers \cite{hu2024routerbench,shnitzer2023large,stripelis2024tensoropera}. For example, UniRoute \cite{jitkrittum2025universal} clusters training data embeddings, represents each LLM based on per-cluster validation errors, and selects the optimal model by balancing the model's performance on the query's cluster and the model cost. CSCR \cite{shirkavand2025cost} focuses on direct model embeddings and selecting the closest model based on cost-aware cosine similarity between the live query embedding and the pre-generated model representations.

\subsection{MPC}

Secure MPC is a cryptographic technology that ensures privacy-preserving data processing. \cite{mohassel2017secureml} decomposes a secret $x \in \mathbb{Z}_{2^l}$ into shares such that $x \equiv \sum_{i=1}^{n} [x]_i \pmod{2^l}$, using a framework that leverages \textit{Linear Homomorphism} to allow participants to locally compute linear combinations $[\alpha x + \beta y]_i = \alpha [x]_i + \beta [y]_i \pmod{2^l}$ without communication~\cite{wagh2019securenn}. \citet{beaver1991efficient, juvekar2018gazelle} address the limitation in multiplication operations and introduce auxiliary primitives (e.g. Beaver Triples), which constitute the primary communication bottleneck. In this paper, we develop PPRoute using secure MPC. In contrast, standard operations performed without MPC are referred to as \textbf{plaintext}.

\subsubsection{Privacy-Preserving Inference for LLM}
Recent works focused on accelerating privacy-preserving Transformer inference through algorithmic approximations and system-level optimizations. To mitigate the latency of non-linear operations (e.g., GeLU, Softmax), frameworks like MPCFormer and SecFormer employ MPC-friendly approximations, such as replacing exponentials with quadratic functions, and use Knowledge Distillation to maintain model accuracy~\cite{li2023mpcformer, luo-etal-2024-secformer}. \citet{pang2024bolt} optimize the underlying cryptographic protocols by integrating a hybrid MPC scheme with an efficient Baby-Step Giant-Step algorithm for homomorphic rotations. Furthermore, addressing the specific challenges of long-context inference, \citet{zeng2025mpcache} introduces MPCache, a mechanism that dynamically discards unimportant tokens from the Key-Value cache to drastically reduce the quadratic overhead of secure attention.

\subsubsection{Secure K-Nearest Neighbor (KNN)}
KNN implementations under MPC introduce prohibitive computational and communication overhead. To these bottlenecks, \citet{li2024secknn} propose \textbf{SecKNN}, a lightweight multi-party framework based on \textit{Function Secret Sharing (FSS)}. By eliminating the dependency on external trusted parties, SecKNN significantly reduces the trust assumptions required for deployment. Crucially, it extends efficient secure computation to diverse metric spaces, introducing novel FSS-based protocols for the Hamming distance and Manhattan distance that enable distance calculation in a single round of interaction. SecKNN achieves inference speeds approximately $50.8\times$ faster. 
The Kona framework \cite{linkona} addresses the cross-term calculation $-2 \sum q_j d_{ij}$ in Euclidean distance by introducing \textit{Euclidean Triples} for pre-computation. This reduces the communication complexity of distance calculation during the online inference phase to zero. Additionally, their \textit{Divide-and-Conquer Bubble Protocol} optimizes  communication rounds of the Top-$K$ selection process from a linear $O(kn)$ to a logarithmic $O(k \log n)$. In addition, Approximate Nearest Neighbor (ANN) algorithms, such as Inverted File Index (IVF) and Hierarchical Navigable Small World (HNSW) graphs, have superior scalability and performance on large-scale datasets \cite{douze2025faiss}.
To address large-scale retrieval, several privacy-preserving ANN algorithm search schemes have been proposed \cite{zhu2025compass, zhou2024pacmann}.

\section{Method}
This section introduces our privacy-preserving LLM routing (PPRoute) framework, which matches the performance of standard embedding-based routing approaches while ensuring user query privacy. PPRoute achieves this by replacing the bottleneck in the transformer encoder with MPC-friendly operations, the multi-stage training algorithm, and the unsorted top-k algorithm in the nearest-neighbor search. The details of PPRoute are shown in \Cref{alg:method}.
\begin{algorithm}[tb]
  \caption{Privacy-Preserving LLMs Routing (PPRoute)}
  \label{alg:method}
  \begin{algorithmic}[1]
    \STATE {\bfseries Initialization:} Generate model representations from the model pool offline. Initialize the transformer encoder, followed by the MLP layer, as the teacher model. Replace the bottleneck functions in the transformer encoder with MPC-friendly operations to get the approx encoder as the student model.\\
    {\bfseries Training in the plaintext:}
    \STATE Freeze the encoder and train MLP layers in the teacher model with embedding-based LLM routing in the plaintext. 
    \STATE Train both the approx encoder and the MLP layers in the student model as follows \cref{eq:loss}.
    \STATE Freeze the approx encoder in the student model and fine-tune the MLP layers only.\\
    {\bfseries Inference in the MPC:}
    \STATE Encrypt the student model and the user query under the MPC. Generate the user query embedding.  
    \STATE Use the ``Unsorted Top-k Algorithm" as shown in \cref{fig:nearest} to finish the nearest neighbor search in the model selection and get the optimal model by balancing semantic relevance and cost constrain.
  \end{algorithmic}
\end{algorithm}

\subsection{MPC-friendly operation} \label{sec3_1}
The query encoder in the LLM routing is a transformer encoder, optionally following an MLP layer. The transformer encoder inference under MPC can be formulated as a 2-Party Computation (2PC), including the input query from the user party and the encoder from the LLM routing provider. Throughout the entire inference process, the LLM routing provider cannot know the user query or the generated query embedding. 

A transformer encoder consists of multi-head attention layers, residual layers, normalization layers, and feed-forward layers. 
Under the MPC, the softmax function in multi-head attention and the Gaussian Error Linear Unit (GeLU) activation are very expensive. Taking a $\text{BERT}_{\text{BASE}}$ model with 12 layers and 512 tokens as an example, the softmax function and GeLU consume 67.8\% and 18.6\% of the total inference time, respectively \cite{li2023mpcformer}. The $\mathrm{Softmax}(x_i) = \frac{\exp(x_i)}{\sum_j \exp(x_j)}$ is slow because the exponential function ($\exp(\cdot)$) is computed through multiple iterative operations (e.g. repeated squaring), which introduces many multiplications. Similarly, GeLU activation is defined as $\mathrm{GeLU}(x) = x \cdot \Phi(x) = \frac{x}{2}\left(1 + \mathrm{erf}\left(\frac{x}{\sqrt{2}}\right)\right)$.  The Gaussian error function ($\mathrm{erf}(\cdot)$) is typically approximated using a high-order Taylor expansion, which involves many multiplication operations and decreases computation accuracy. Therefore, the PPRoute uses MPC-friendly operation to replace bottleneck functions in the transformer encoder under MPC. 

\textbf{Softmax approximation.} Since the exponential function is unacceptable and we use $\mathrm{ReLU}(x) = max(0, x)$ to replace it \cite{mohassel2017secureml, li2023mpcformer}. We call the softmax function approximation ``2ReLU” and reformulate it as:
\begin{equation}
\text{softmax}(x) \approx \mathrm{ReLU}(x) / \sum \mathrm{ReLU}(x) 
\end{equation}
\textbf{GeLU approximation.} 
The Gaussian error function is computationally expensive and suffers from reduced numerical accuracy under MPC. Since $\mathrm{ReLU}(x)$ is simple, efficient, and widely adopted in machine learning models, PPRoute uses $\mathrm{ReLU}(x)$ to approximate GeLU activation. In addition, $\mathrm{ReLU}(x)$ can be applied directly in MLP layers if LLM routing includes MLP layers.  

\textbf{The approx encoder.} We build the approx encoder using ``2ReLU" and ``ReLU" to replace softmax and ``ReLU", separately, as Line 1 in \cref{alg:method}.


\begin{figure}[t]
  \begin{center}
    \centerline{\includegraphics[width=0.7\columnwidth]{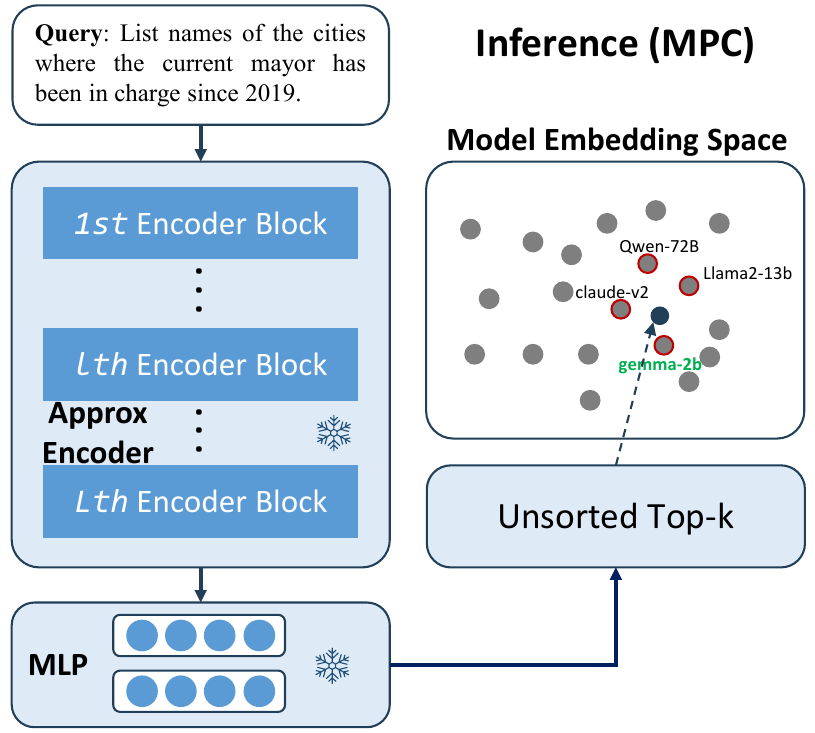}}
    \caption{Overview of Privacy-Preserving LLMs Routing (PPRoute) Inference. 1) Generate query embedding; 2) Use Unsorted-topk to do retrieval or clustering.
    }
    \label{pprouteinference}
  \end{center}
\end{figure}
\subsection{Multi-stage Training Algorithm,}
The use of MPC-friendly operations directly changes the architecture of the pre-trained transformer encoder, which damages model performance. In this subsection, we use a multi-stage training algorithm to train the approximated encoder in the plaintext to make it have similar performance as the original encoder. Here, we denote the approximated encoder as the student model and the original encoder as the teacher model. Both models include MLP layers after the encoder for further projections.

Initially, the teacher model's encoder is frozen and the teacher model’s MLP layer is trainable. 
Second, we use a hybrid training method by combining routing training $\mathcal{L}_{routing}$ and model distillation $\mathcal{L}_{Distill}$, where $\mathcal{M}_t$ denotes the teacher model and $\mathcal{M}_s$ denotes the student model with an approx encoder. Finally, we freeze the student model's encoder and optionally fine-tune the MLP layers in the student model to improve the model's performance further. 

\begin{equation}
\label{eq:loss}
\min_{\Theta} \;\;
\mathcal{L}(\mathcal{M}_s)
\;:=\;
\mathcal{L}_{\text{routing}}(\mathcal{M}_s)
\;+\;
\beta \, \mathcal{L}_{\text{distill}}(\mathcal{M}_t \,\|\, \mathcal{M}_s),
\end{equation}
where $\Theta$ consists of trainable parameters of the whole student model and the MLP layers of the teacher model: $\Theta = [\theta_{\text{Enc}}^s, \theta_{\text{MLP}}^s, \theta_{\text{MLP}}^t ]$

Equivalently, the objective in Eq.~(\ref{eq:loss}) can be written as the trust-region constrained optimization problem in Eq.~(\ref{eq:trust_region}).
\begin{equation}
\label{eq:trust_region}
\min_{\Theta} \;\;
\mathcal{L}_{\text{routing}}(\mathcal{M}_s)
\quad \text{s.t.} \quad
\mathcal{L}_{\text{distill}}(\mathcal{M}_t \,\|\, \mathcal{M}_s)
\;\le\;
\epsilon ,
\end{equation}
where we minimize the routing loss while constraining the student to remain close to the teacher under the distillation loss. Here, $\epsilon$ defines the radius of the trust-region and controls the maximum allowable deviation from the teacher.

It should be noted that model training is completed in the plaintext setting, and MPC is only applied in the inference stage.

\subsection{Nearest Neighbor Search}
In this subsection, we introduce the unsorted top-k algorithm used in PPRoute, which is MPC-friendly and tailored for small-scale retrieval.

In the plaintext embedding-based LLM routing protocol, the model selection logic is usually designed as multiple rounds of nearest neighbor search. In UniRoute, the input query is assigned to a cluster center. Then, it retrieves the model from the model pool based on the sum of the error score and a cost-adjustment factor. In the CSCR, it first retrieves the top-$k$ models based strictly on the embedding similarity between the model and the query, and then selects the candidate with cost-aware cosine similarity as \cref{eq:topk}. This strategy is essential to prevent the selection of low-quality nodes at a very low cost.

Although this workflow is trivial to implement in plaintext, multiple rounds of KNN in the context of MPC are computationally prohibitive. Performing a full sort or global search over all secret-shared neighbors incurs high communication overhead as a result of the extensive use of secure comparison operations. Furthermore, simply applying current state-of-the-art (SOTA) secure top-$k$ protocols to identify the candidate set remains a performance bottleneck for real-time inference. Therefore, we introduce a specialized MPC-friendly sorting algorithm, unsorted top-$k$, which is specifically designed for small datasets in MPC.
\begin{figure}[htbp]
    \centering
    \includegraphics[width=0.48\textwidth]{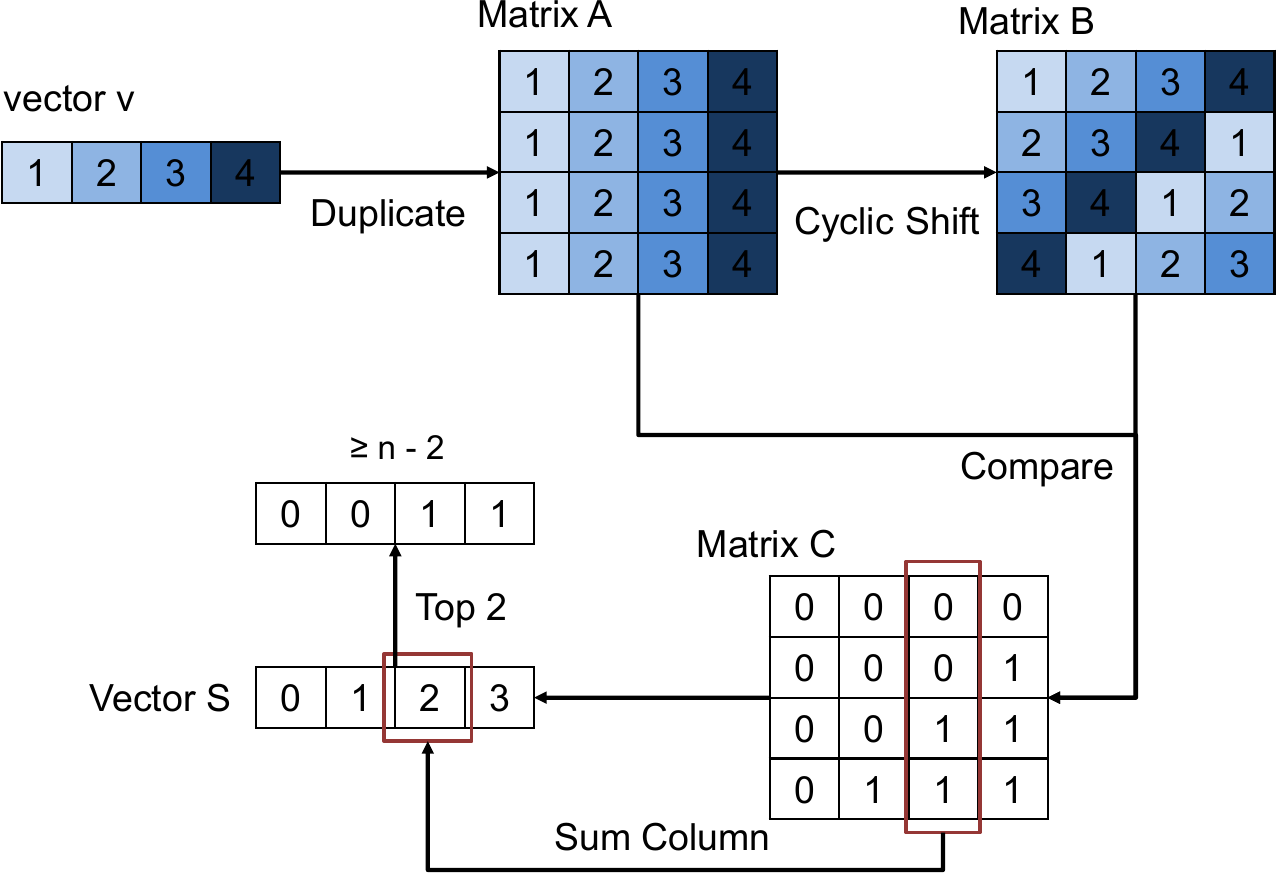}
    \caption{The Unsorted Top-k Algorithm Workflow} 
    \label{fig:nearest} 
\end{figure}

The ``Unsorted Top-k'' algorithm restructures the selection problem into highly parallelizable matrix operations to optimize Multi-Party Computation (MPC) performance, shown in \Cref{fig:nearest}. The top K sorting begins with the construction of a comparison matrix, where the secret-shared input vector $V$ is replicated and cyclically shifted to form matrices $A$ and $B$
. A binary comparison matrix $C$ is then generated through element-wise comparisons, formally defined as $C_{i,j} = \mathbbm{1}(A_{i,j} > B_{i,j})$. Although this step entails a total Computation complexity of $O(n^2)$, the bottleneck in the MPC is usually communication complexity, the independence of each matrix entry allows parallel execution and the size of the model pool is limited. Consequently, communication complexity is bounded by the constant depth of a single secure comparison, $O(1)$, effectively mitigating the latency overhead typically associated with iterative comparison protocols.

Following the comparison phase, the algorithm derives a ranking score vector $S$ through local aggregation. The rank of each element is determined by summing the columns of the comparison matrix, expressed as $S_j = \sum_{i=1}^n C_{i,j}$. Because this summation involves only local additions of secret shares, it incurs zero communication cost. The final selection of the top-$k$ elements is achieved by applying a threshold function to these scores against a public constant. Specifically, a mask vector $M$ is calculated where $M_j = 1$ if $S_j \geq n-k$. This formulation allows the entire protocol to complete in constant rounds, $O(1)$, ensuring that for small $n$, the computational bottleneck shifts entirely from network latency to bandwidth. The unsorted top-K algorithm is well-suited for LLM routing retrieval, as the size of the model pool is limited.

\subsection{Application on the CSCR}
PPRoute could be used in many embedding-based LLM routing since they mostly include two components: encoder inference and nearest neighbor search. We apply PProute on the CSCR \cite{shirkavand2025cost} as shown in \cref{pproute}. The query encoder in CSCR uses a transformer encoder followed by a two-layer neural network. CSCR generated model representation with Logit-footprint Descriptors for open LLMs and perplexity fingerprints for Black-Box LLMs, separately. It proposes Cost-Spectrum InfoNCE to train the routing encoder. In the routing inference, the user query is converted to an embedding by the encoder, and two steps of nearest neighbor search are conducted: 1) the routing retrieves k most similar LLMs to the user query embedding from the model pool; 2) then the routing selects the optimal model based on cost-aware cosine similarity as below:
\begin{equation} \label{eq:topk}
\hat{r}(\mathbf{x})=\underset{h \in \operatorname{Top}_k(\mathbf{x})}{\arg \max }\left[\cos \left\langle \mathbf{e}_{\text{query}}, \mathbf{d}_h\right\rangle-\lambda c(h)\right]
\end{equation}
where $h \in \operatorname{Top}_k(\mathbf{x})$ retrieves the k models most similar to the user query, $\cos \langle\rangle$ means cosine similarity, $\mathbf{e}_{\text{query}}$ denotes the embedding of the user query, $\mathbf{d}_h$ is the model representation, $\lambda$ is the cost weight, and $c(h)$ means the cost function of the model candidate. 

To apply PPRoute to the CSCR, 1) We replace bottleneck functions with MPC-friendly operations to serve as the student model. 2) Following the CSCR approach, we utilize Cost-Spectrum InfoNCE as $\mathcal{L}_{\text{routing}}$ to train the MLP layers within the teacher model. Similarly, the student model is trained according to the loss function defined in \cref{eq:loss} with Cost-Spectrum InfoNCE. 3) During inference, we use MPC to encrypt both the student model and the user query, ensuring that the resulting query embedding remains private. 4) PPRoute executes an Unsorted-Topk operation to identify the $k$ most similar models, followed by a second Unsorted-Topk pass to select the optimal candidate as \cref{eq:unsorttopk}.
\begin{equation} \label{eq:unsorttopk}
\hat{r}(\mathbf{x})=\underset{h \in \operatorname{Unsorted-Top}_k(\mathbf{x})}{\arg \max }\left[\cos \left\langle \mathbf{e}_{\text{query}}, \mathbf{d}_h\right\rangle-\lambda c(h)\right]
\end{equation}

\subsection{Application on the UniRoute}
Similarly, PPRoute can be applied to UniRoute. The UniRoute process is implemented as follows: 1) UniRoute first clusters training data embeddings into $K$ distinct groups and assigns each prompt from the validation set to its corresponding clustering center; 
2) Model representations are generated offline and consist of a vector of length $K$, where each floating value represents the mean performance score for that specific cluster on the validation set. 
During inference, the user query is processed through a transformer encoder and the query embedding is assigned to a clustering center based on embedding similarity. The best model is selected by minimizing the sum of the error score and a cost-adjustment factor, defined as: $\text{Score} + \lambda \cdot \text{Model cost}$. PProute can optimize the query encoder under MPC and use unsorted topk to do clustering and search for the best candidate.

\begin{table}[t]
    \centering
    \caption{Time Comparison of CSCR in MPC without and with PPRoute optimization}
    \label{cscrtime}
    \resizebox{\columnwidth}{!}{
        \begin{tabular}{lccc}
            \toprule
            \textbf{Dataset} & \textbf{Baseline (s)} & \textbf{Optimized (s)} & \textbf{Speedup} \\
            \midrule
            embedllm    & 199.47 & 9.38 & 21.26$\times$ \\
            mixinstruct & 183.44 & 9.12 & 20.11$\times$ \\
            routerbench & 185.73  & 9.11 & 20.28$\times$ \\
            \bottomrule
        \end{tabular}
    }
\end{table}

\begin{table}[t]
    \centering
    \caption{Time Comparison of UniRoute in MPC without and with PPRoute optimization}
    \label{umrtime}
    \resizebox{\columnwidth}{!}{
        \begin{tabular}{lccc}
            \toprule
            \textbf{Dataset} & \textbf{Baseline (s)} & \textbf{Optimized (s)} & \textbf{Speedup} \\
            \midrule
            embedllm    & 190.29 & 9.90 & 19.22$\times$ \\
            mixinstruct & 188.09 & 8.07 & 23.31$\times$ \\
            routerbench & 186.35 & 8.21 & 22.70$\times$ \\
            \bottomrule
        \end{tabular}
    }
\end{table}

\section{Experiments}
We design the PPRoute framework to be compatible with embedding-based LLM routing under MPC. Thus, we evaluate PPRoute under (1) Time cost, (2) Routing quality. We evaluate PPRoute applied to CSCR in \cref{sec:4_1}, \cref{sec:4_2}, and \cref{sec:4_3}, and we present the application on UniRoute in \cref{sec:4_4}.

\begin{table}[t]
  \caption{The communication time statistics for the transformer encoder with MPC-friendly operation approximations.}
  \label{MPC-operation}
  \begin{center}
    \begin{small}
      \begin{sc}
        \begin{tabular}{lcccr}
          \toprule
          FeedFoward & Multi-Head Attention  & Time (s)   \\
          \midrule
          GeLU  & Softmax &  189.32 \\ 
          GeLU  & 2ReLU &  8.35 \\ 
          ReLU  & 2ReLU &  7.38 \\ 
          \bottomrule
        \end{tabular}
      \end{sc}
    \end{small}
  \end{center}
  \vskip -0.1in
\end{table}

\subsection{Settings} \label{sec:4_1}
\textbf{Datasets.} We evaluate our router on three benchmark datasets: EmbedLLM \cite{zhuang2024embedllm}, MixInstruct \cite{jiang2023llm}, and RouterBench \cite{hu2024routerbench} with one user query. To apply PProute to CSCR \cite{shirkavand2025cost}, we adopt the experimental setup from CSCR and sampled 192 probes from the validation sets of EmbedLLM and MixInstruct. For each model, we generated logit-based descriptors by capturing the top-$K$ tokens ($K = 256$) on a prediction horizon of $T = 10$. For RouterBench, we sampled 192 probes from the training split, using GPT-2 \cite{radford2019language} to compute perplexity-based descriptors.

\begin{table*}[t]
    \centering
    \resizebox{\textwidth}{!}{%
    \begin{threeparttable}
        \caption{Performance Comparison of Different Retrieval Methods in embedllm dataset}
        \label{tab:time_comparison}
        \begin{tabular}{lrrrrrl}
            \toprule
            \textbf{Method} & \textbf{Comm. Vol} & \textbf{Comm.} & \textbf{Comm. Time} & \textbf{Comp.} & \textbf{Total} & \textbf{Type} \\
             & (Bytes) & \textbf{Round} & (20 GBps) & \textbf{Time (s)} & \textbf{Time (s)} & \\
            \midrule
            
            Baseline~\cite{knott2021crypten} & 24,962,400 & 130 & $ 0.0967 \pm 0.0040$ & $0.5039 \pm 0.0157$ & $0.6006 \pm 0.0183$ & Top-$k$ \\
            \addlinespace
            
            ANN~\cite{malkov2018efficient} $^{*}$ & 14,977,440 & 78 &  $0.0620 \pm 0.0015$  &  $0.2823 \pm 0.0075$ & $ 0.3443 \pm 0.0072$ & Top-$k$ \\
            \addlinespace
            
            Bitonic Sort~\cite{batcher1968sorting} & \makecell[r]{1,234,832} & \makecell[r]{28,682} & $ 8.3137 \pm 0.2845$ & $ 35.1922\pm0.1603$ & $43.5059 \pm 0.3435$ & Sorting \\
            \addlinespace
            
            DB Bubble~\cite{linkona} & \makecell[r]{299,968} & \makecell[r]{6,976} & $2.0052 \pm 0.0591$ & $8.6293 \pm  0.0711$ & $10.6345 \pm 0.1054$ & Sorting \\
            \addlinespace
            
            Brute Force~\cite{shirkavand2025cost} & 290,464 & 6,756 & $1.9422 \pm 0.0576$ &  $ 8.0884 \pm 0.0768$ & $10.0650 \pm 0.0976$ & Top-$k$ \\
            
            \midrule 
            
            \textbf{Unsorted Top-$k$ (Ours)} & \textbf{9,984,960} & \textbf{52} & \textbf{0.0472 $\pm$ 0.0014} & \textbf{0.2008 $\pm$ 0.0112} & \textbf{0.2480 $\pm$ 0.0116} & Top-$k$ \\
            
            \bottomrule
        \end{tabular}
        \begin{tablenotes}
            \footnotesize
            \item[*] 1. This method requires \textbf{Oblivious RAM (ORAM)} for security implementation.
            \item 2. Results are reported as $\text{Mean} \pm \text{Standard Deviation}$ over 10 runs.
            \item 3. Comm. denotes Communication; Comp. denotes Computation.
        \end{tablenotes}
    \end{threeparttable}
    }
\end{table*}

\textbf{Implementation.} We follow the MPC configuration in \cite{li2023mpcformer} and the time breakdown under MPC was measured via CrypTen, a framework that extends PyTorch to support privacy-preserving machine learning. It implements secret sharing with the semi-honest adversaries assumption \cite{knott2021crypten}. 
We utilize a \texttt{sentence-transformers/all-MiniLM-L6-v2} model as the embedding encoder across all experiments, with the sequence length 256. The MLP layers are implemented as a two-layer neural network with the ReLU activation function. We train the LLM routing on the training split of each dataset. For the cost-spectrum loss \cite{shirkavand2025cost}, we set the number of cost bands as $K = 5$. Hyperparameters, including band-specific temperatures $\alpha$ and $\tau_{\min}$ and the negative cost penalty $\gamma$ are tuned via grid search in the range [0.20, 0.25, 0.30, 0.35, 0.40, 0.45], [0.01, 0.05, 0.1, 0.15, 0.2, 0.25] and [0.15, 0.2, 0.25, 0.3, 0.35, 0.4], separately.

\textbf{Evaluation.} We evaluate the time cost in \cref{sec:4_2}. We evaluate the routing quality in \cref{sec:4_3} using deferral curves \cite{jitkrittum2025universal,shirkavand2025cost}, which visualize the trade-off between the average response quality and the total inference cost. These curves are generated by sweeping the routing penalty parameter $\lambda$. We use 1) Area Under the Deferral Curve (AUDC): Measures the overall efficiency of the router across all cost-performance trade-offs; 2) Peak Accuracy: The maximum performance achieved by the system regardless of cost; 3)Query-Normalized Cost (QNC): The minimum relative cost required to match the performance of the most accurate Large Language Model (LLM) in the evaluation set. The inference cost varies by dataset \cite{shirkavand2025cost}: The Cost in EmbedLLM and MixInstruct is defined by the number of model parameters, serving as a proxy for computational overhead and latency. The Cost in RouterBench is calculated using actual API invocation fees (USD). 

\subsection{Time Cost Comparison} \label{sec:4_2}
In this section, we validate the effectiveness of our PProute at the cost in time. We apply PPRoute on CSCR and UniRoute and compare it with na\"ve implementation of encoder and brute-force retrieval under MPC. As shown in \Cref{cscrtime} and \Cref{umrtime}, PPRoute achieves consistent speedup across different benchmark datasets.

\begin{table*}[!h]
\centering
\caption{Comparison of routing methods on EmbedLLM, Mix-Instruct, and RouterBench.
Higher AUDC and Peak are better, while lower QNC is better.}
\setlength{\tabcolsep}{6pt}
\renewcommand{\arraystretch}{1.2}
\resizebox{0.865\textwidth}{!}{
\begin{tabular}{lccccccccc}
\toprule
& \multicolumn{3}{c}{\textbf{EmbedLLM}} 
& \multicolumn{3}{c}{\textbf{Mix-Instruct}} 
& \multicolumn{3}{c}{\textbf{RouterBench}} \\
\cmidrule(lr){2-4} \cmidrule(lr){5-7} \cmidrule(lr){8-10}
\textbf{Router} & AUDC$\uparrow$ & QNC$\downarrow$ & Peak$\uparrow$
& AUDC$\uparrow$ & QNC$\downarrow$ & Peak$\uparrow$
& AUDC$\uparrow$ & QNC$\downarrow$ & Peak$\uparrow$ \\
\midrule
Oracle (upper bound)
& 0.7755 & 8.5 & 0.979
& 0.077 & 32.3 & 0.081
& 0.812 & 0.290 & 0.910 \\
Random
& 0.3710 & 50.3 & 0.417
& 0.039 & 18.5 & 0.043
& 0.4840 & 0.575 & 0.531
\\
\cdashline{1-10}
\multicolumn{10}{c}{CSCR \cite{shirkavand2025cost}} 
\\
\cdashline{1-10}
Original
& 0.539 & 44.4 & 0.568
& 0.048 & 34.1   & 0.050
& 0.6843  & 1.660 &  0.794 \\
\textbf{PPRoute (Ours)}
& 0.538 & 45.3 & 0.564
& 0.048  & 34.8 & 0.050
& 0.6844  & 1.660 & 0.794 \\
\cdashline{1-10}
\multicolumn{10}{c}{UniRoute \cite{jitkrittum2025universal}} 
\\
\cdashline{1-10}
Original 
& 0.515 & 60.21 &  0.562
& 0.047 & 36.7 & 0.050 
& 0.6661  & 1.659 & 0.794 \\
\textbf{PPRoute (Ours)}
& 0.5036 & 64.0 &  0.560
& 0.047 & 37.6 & 0.050
& 0.6662 & 1.66 & 0.794 \\
\bottomrule
\end{tabular}
}
\label{quality_result}
\end{table*}

\begin{figure*}[!h]
\centering
\subfigure[ EmbedLLM ]{
\hspace{0pt}
\includegraphics[width=.31\textwidth]{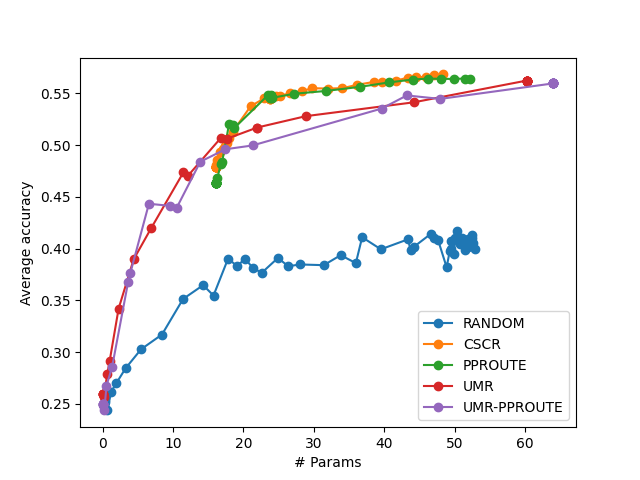}
}
\subfigure[Mix-Instruct]{
\hspace{0pt}
\includegraphics[width=.31\textwidth]{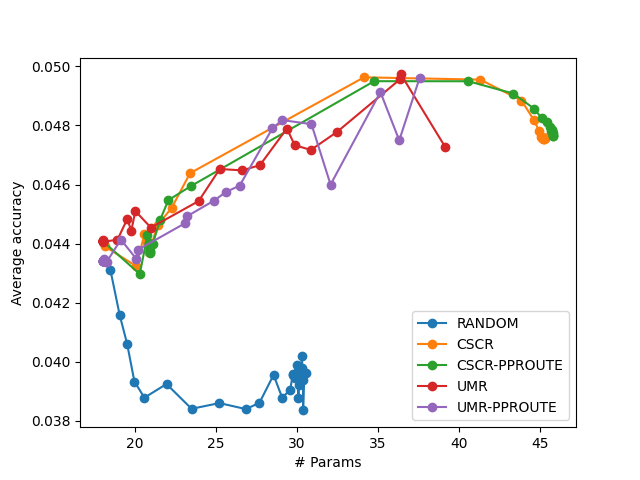}
}
\subfigure[RouterBench]{
\hspace{0pt}
\includegraphics[width=.31\textwidth]{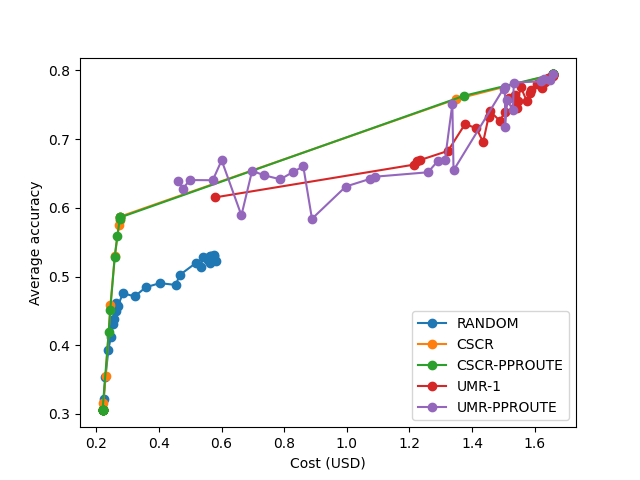}
}
\caption{Accuracy–cost/size deferral curves on three datasets}
\label{fig:1}
\end{figure*}

\subsubsection{Routing Encoder Inference}
We evaluate the runtime under the MPC setting. We replace softmax and GeLU in the encoder with ``2ReLU" and ReLU, separately. We validate the approximation operations in \cref{MPC-operation}, which shows that our methods can significantly reduce the time consumption of the LLM routing transformer encoder in MPC inference.

\subsubsection{Nearest Neighbors search}
In this part, we present the performance of the unsorted Top-k Algorithm in PPRoute. It optimized sorting time cost, the bottleneck in the K-Nearest Neighbors search. 

We use CSCR as an application. It uses query embedding to search the top-4 models from the model pools and then get the optimal model from 4 candidates based on the cost-aware cosine similarity. CSCR used the brute force algorithm in FAISS~\cite{douze2025faiss} to perform a $k$-NN lookup.

We compare our proposed Unsorted Top-$k$ algorithm against a comprehensive set of baselines to evaluate its efficiency. These include: (1) \textbf{Baseline}~\cite{knott2021crypten}, which leverages CrypTen's built-in \texttt{max} function to iteratively perform the Top-$1$ operation four times followed by a final aggregation; (2) \textbf{ANN}~\cite{malkov2018efficient}, a simplified approximation based on Hierarchical Navigable Small World (HNSW) graphs, chosen because standard HNSW incurs excessive overhead when constructing multi-layer structures for small datasets (noting that this method requires ORAM for security); (3) \textbf{Bitonic Sort}~\cite{batcher1968sorting}, a classical oblivious sorting network; (4) \textbf{DB Bubble}~\cite{linkona} (Kona), an optimized oblivious bubble sort algorithm; and (5) \textbf{Brute Force}~\cite{shirkavand2025cost}, which corresponds to the plaintext retrieval strategy employed in CSCR, exhaustively enumerating all possibilities to identify the maximum value.

\Cref{tab:time_comparison} presents performance metrics, including communication volume, rounds, and exact execution time. Our Unsorted Top-$k$ consistently outperforms all relevant baselines, achieving the lowest time cost of \textbf{0.2518s}. Specifically, compared to the \textbf{Baseline}, our method reduces both communication volume (from 24.9 MB to 9.9 MB) and communication rounds (from 130 to 52) by approximately 60\%, resulting in a $2.5\times$ speedup. While sorting-based methods like \textbf{Bitonic Sort} and \textbf{DB Bubble}, as well as the CSCR-aligned \textbf{Brute Force}, achieve lower communication volumes, they are severely bottlenecked by high round complexity (thousands of rounds) and computation overhead, resulting in execution times exceeding 10 seconds. Finally, our method also exceeds \textbf{ANN} (0.3786s) in efficiency without requiring complex ORAM infrastructure.


\subsection{Quality Comparison with CSCR} \label{sec:4_3}
In this section, we apply our PPRoute framework to CSCR to provide privacy protection and compare its performance with the plaintext setting. In addition, we also include two baselines: Random, which selects models uniformly to simulate na\"ive routing, and Oracle, which consistently chooses the most accurate model at the minimum cost to establish a theoretical performance ceiling. \Cref{quality_result} and \Cref{fig:1} show that though PPRoute replaced the bottleneck function in the transformer model, it still gets similar results under AUDC, QNC and Peak metrics across different datasets. In particular, PPRoute successfully balances performance and inference cost while providing user privacy protection.


\subsection{Quality Comparison with UniRoute} \label{sec:4_4}
To demonstrate the generalization capabilities of our framework, we also apply our PPRoute framework to another embedding-based routing, UniRoute \cite{jitkrittum2025universal}. \Cref{quality_result} and \Cref{fig:1} show that PPRoute maintains consistent performance across various datasets compared to UniRoute. This highlights that PPRoute can be widely applied to various embedding-based LLM routing and reduce privacy risk.

\section{Conclusion}
Although LLM routing is essential for balancing model performance and cost-efficiency, it introduces significant privacy risks that have remained largely unaddressed. To bridge this gap without the prohibitive latency typically associated with cryptographic methods, we propose PPRoute, a privacy-preserving framework for embedding-based routing. By implementing targeted acceleration strategies for both secure encoder inference and nearest neighbor search, PPRoute effectively mitigates the computational overhead of Secure Multi-Party Computation (MPC). Ultimately, PPRoute ensures that users can benefit from the economic and functional advantages of multi-model routing while maintaining robust, end-to-end data privacy. Applications on the CSCR and UniRoute validate its effectiveness.

\section*{Impact Statement}
This paper aims to advance machine learning by improving privacy in large language model (LLM) routing. The proposed method helps protect user queries by allowing routing decisions to be made without revealing the content of the data. This can benefit real-world systems, especially in areas such as healthcare, finance, and enterprise applications where data privacy is important. Overall, this work supports the development of safer and more privacy-aware machine learning systems, and we do not foresee any negative societal impacts

\bibliography{example_paper}
\bibliographystyle{icml2026}




\end{document}